\documentclass[reprint,aps,prl,amsmath,amssymb,superscriptaddress,showpacs]{revtex4-1}
\usepackage{tikz,graphicx,url,float}
\usetikzlibrary{arrows}

\usepackage{amsthm}

\begin{document}

\title{Tropical approximation to finish time of activity networks}

\author{Alexei Vazquez}
\email{alexei@nodeslinks.com}
\affiliation{Nodes \& Links}

\date{\today}

\begin{abstract}
We breakdown complex projects into activities and their logical dependencies. We estimate the
project finish time based on the activity durations and relations. However, adverse events trigger delay cascades shifting the finish time. Here I derive a tropical algebraic equation for the finish time of activity networks, encapsulating the principle of linear superposition of exogenous perturbations in the tropical sense. From the tropical algebraic equation I derive the finish time distribution with explicit reference to the distribution of exogenous delays and the network topology and geometry.
\end{abstract}

\maketitle

In recent years we have experienced a significant advance in our understanding of complex networks and of processes running on them \cite{albert02,dorogotsev08,pastor15}. Yet, we still lack an understanding of the interplay between network topology and dynamics in the context of activity networks. There is a vast literature on Monte Carlo simulations of activity networks under uncertainty and risk events \cite{acebes15,moret16}. The critical path method has been used to aggregate perturbations and derive an analytical approximation to the project end date distribution \cite{van13}. However, activity networks are characterised by a complex topology \cite{braha04,braha07,pozzana21} and activity delays exhibit a high frequency of extreme events \cite{pozzana21,araujo21,park21}. In that context the key conditions for the critical path method,  the existence of a dominant path from the project start to its end and the central limit theorem, do not apply. Here I obtain a tropical algebra approximation to the project end date.

Let $P(V,E,\vec{d})$ be a project schedule with a set of activities $V$, a set of activity relations $E$ and a vector of activity durations $\vec{d}$. The project start and end are represented by the activities $i=1$ and $i=n$, respectively, of duration zero. An arc $i\rightarrow j\in E$ indicates that $i$ must finish before $j$ starts. I will denote the set of activity predecessors of $i$ by $I_i=\{j|j\rightarrow i\in E\}$. Logical consistence implies that the project network is a directed acyclic graph.  The nodes in a directed acyclic graph can be ordered topologically such that if $i\rightarrow j$ then $i$ is before $j$ in the topological order, for all $i\rightarrow j\in E$.

The earliest an activity can finish is determined by the recursive relation
\begin{equation}
x_i = d_i + \max_{j\in I_i} x_j\ ,
\label{forward} 
\end{equation}
with the boundary condition $x_1=s$, where $s$ is the start date. We calculate $\vec{x}$ with a forward pass of  Eq. (\ref{forward}) along the topological order. Next we do backward propagation to calculate the latest an activity can finish without altering the project end date
\begin{equation}
y_i = \min_{j | i \in I_j} \left(y_j - d_j\right)\ ,
\label{backward} 
\end{equation}
with the boundary condition $y_n=x_n$. $\vec{x}$ and $\vec{y}$ set the early and late start dates for every activity. If $i\in I_j$ then 
\begin{equation}
w_{ji} = x_j - x_i\ ,
\label{w}
\end{equation}
is the free float, the maximum delay at $i$ that does not shift the early finish date of $j$ \cite{van13}. In turn we can determine the amount of delay tolerated at a given activity without causing a shift in the project end date. This is the total float and it is calculated as
\begin{equation}
T_{ni} = x_i-y_i\ .
\label{Tn} 
\end{equation}
The subscript $n$ emphasises that $T_{ni}$ is the total float from activity $i$ to the project end. This will be generalised below to any pair of activities. I have adopted a negative sign for free floats and total floats to indicate delay subtraction.

In practice exogenous factors delay the activity starts or increase activity durations \cite{park21}. If the finish delay of an activity exceeds the free float of any successor, then it will cause their delay as well, starting a delay cascade \cite{braha05,ellinas19}. Let $\vec{h}$ represent the vector of activity delays caused by exogenous factors and $\vec{z}$ the vector of activity delays after the propagation of the exogenous delays. The delays propagate via the recursive equation
\begin{equation}
z_i = f_i\left\{\max_{j\in I_i} \left[\max(0, w_{ij} + z_j)\right] , h_i\right\}\ .
\label{forward_delay} 
\end{equation}
The term $\max(0,w_{ij}+z_j)$ indicates that delays are passed if they exceed the free float between the activities. Then we take the maximum delay from $j\in I_i$. The function $f_i(x,y)$ merges the delays coming from predecessors (endogenous) and exogenous sources. If exogenous delays act on activity starts, then $f_i(x,y)=\max(x,y)$. If exogenous delays act on activity durations, then $f_i(x,y)=x+y$. Even a unique contingency such as an adverse weather event can act differently depending on where it falls in the calendar relative to an activity. If it falls before the activity start, it competes with delays from predecessors as a cause of delay for the activity start. If the adverse weather event happens while the activity is ongoing it will delay its finish date on top of any delay coming from predecessor activities. In general,
\begin{equation}
\max(x,y)\leq f_i(x,y)\leq x+y\ .
\label{f_bounds}
\end{equation}
To be precise we would specify $f_i(x,y)$ for every combination of activity and adverse event. Yet, when $x$ and $y$ follow sub-exponential distributions $\max(x,y)\approx x+y$. If that is the case then $f_i(x,y)\approx \max(x,y)\approx x+y$. 

Sub-exponential distributions cannot be bounded by any exponential function $f(x) = e^{-\alpha x}$  when $x\rightarrow\infty$ \cite{foss13}. If we take two random variables $x$ and $y$ generated from the same sub-exponential distribution then ${\rm Pr}[x+y>z]\sim {\rm Pr}[\max(x,y)>z]$ when $z\rightarrow\infty$. The only way that $x+y$ is larger than $z$ is for $x$ or $y$ to be greater than $z$. We cannot take this result as granted when we propagate delays in the activity network. The errors may accumulate when the delay propagation takes several steps and free floats act as delay sinks. In the following I investigate the difference in using $f=\max$ or $f={\rm sum}$ by means of numerical simulations.

We need a model to generate project networks, a model to generate activity durations and a model to generate exogenous delays. We will generate project networks using the duplication-split model \cite{vazquez22}. In this model, we start with two activities representing the project start and end, with an arc from start to end. Then, at each discrete step, we select an activity  $i$ with uniform probability across all current activities and create a new activity $j$. With probability $q$, $j$ is a duplicate of $i$, inheriting all the incoming and outgoing relations of $i$. Otherwise, $i$ transfers all the outgoing arcs to $j$ and an arc from $i$ to $j$ is created. We call $q$ the duplicate rate.  It is demonstrated that both the in-degree and out-degree distribution has a power law tail with exponent $1/q$ and the network diameter grows as $n^{1-q}$ \cite{vazquez22}. As we tune $q$ from 0 to 1 we change from networks that are quasi-linear and have narrow degree distributions to networks with wide degree distributions and multiple parallel paths.

The simulations proceed as follow. The inputs are the number of activities $n$, the duplication rate $q$, the distributions of activity durations $p(d)$ and the exogenous delay distribution $p(h)$. We focus on sub-exponentially distributed exogenous delays, so I choose the log-normal $p(h)\sim e^{-[log(h/\mu)]^2/2\sigma^2}/h$ with $\mu=1$. First, we generate a duplication-split network  with parameters $(n,q)$. Second, we assign durations to activities from $p(d)$ and run Eqs. (\ref{forward})-(\ref{w}) to generate the free floats $w$. Third, we generate random delays from $p(h)$ and run Eq. (\ref{forward_delay}) with a pre-defined $f(x,y)$ to calculate $z_n$, the project end delay. We repeat this third step to generate $z_n$ samples and report the smallest $z_n$ that is larger than 80\% of all samples. This quantity is commonly used in project management and it is named the project $p80$. Fourth, for each parameter set $\{n,q,p(d),p(h),f(x,y)\}_i$ we generate multiple $p80$ values $\pi(f)_i$ by sampling over realisations of the network and activity durations. We do that for $f=\max$ and $f={\rm sum}$. Finally, we calculate the slope through the origin
\begin{equation}
S = \frac{ \sum_i \pi(f={\max})_i \pi(f={\rm sum})_i }{ \sum_i [\pi(f={\max})_i]^2 }
\label{slope}
\end{equation}
Since $x+y\geq\max(x,y)$ then $S\geq1$.

\begin{figure}[t]
\includegraphics[width=3.3in]{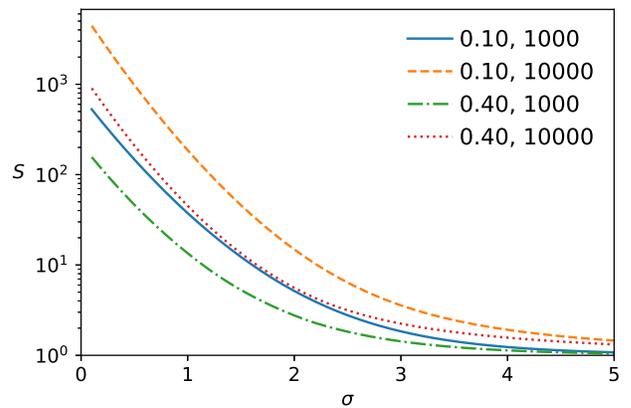}
\caption{Slope between the calculated p80s using $f={\rm sum}$ vs using $f={\max}$, for $\vec{d}=\vec{0}$ and the $(q,n)$ indicated in the legend.}
\label{fig1}
\end{figure}

\begin{figure}[t]
\includegraphics[width=3.3in]{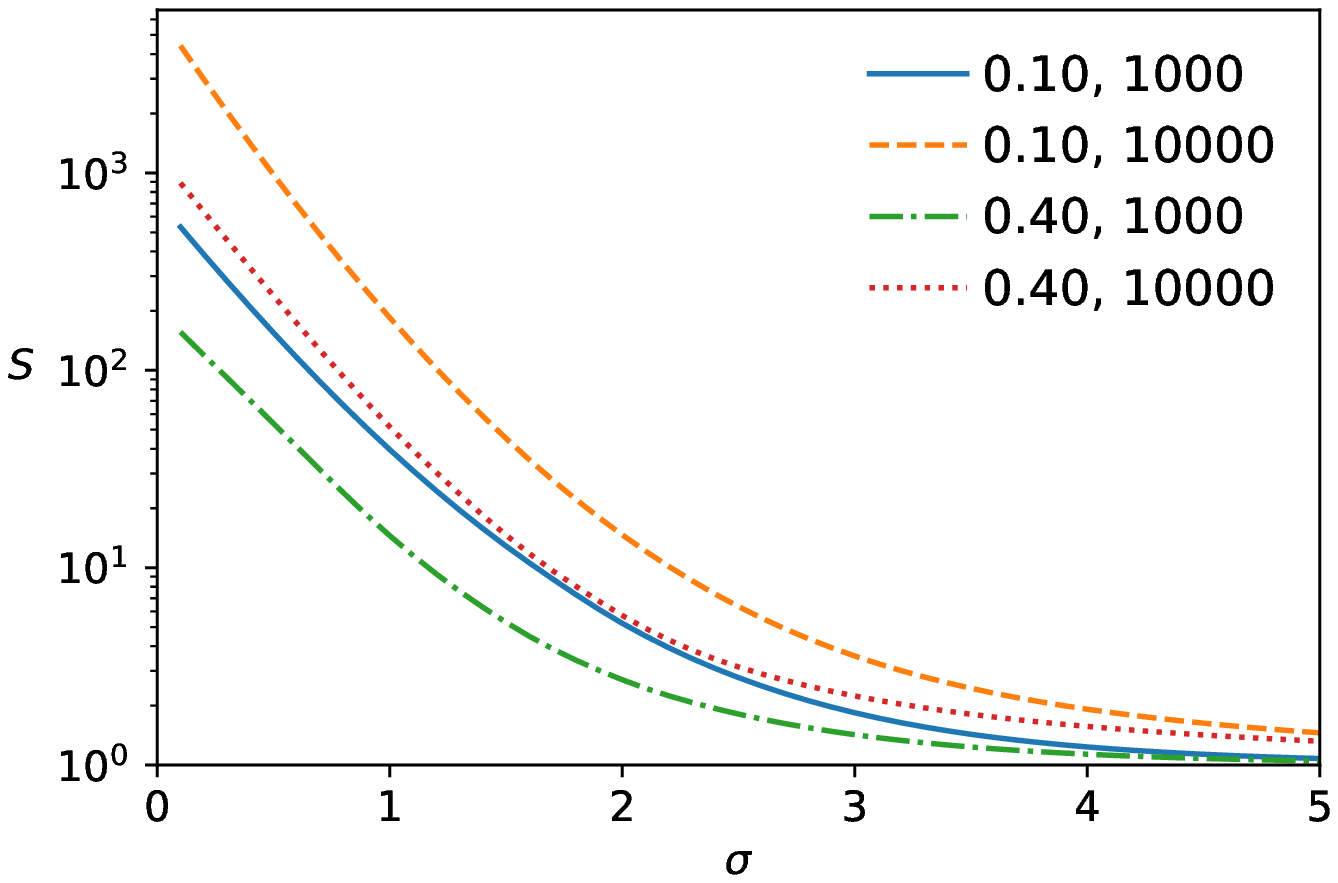}
\caption{Slope between the calculated p80s using $f={\rm sum}$ vs using $f={\max}$, for $\sigma_1=1$ and the $(q,n)$ indicated in the legend. }
\label{fig2}
\end{figure}

\begin{figure}[t]
\includegraphics[width=3.3in]{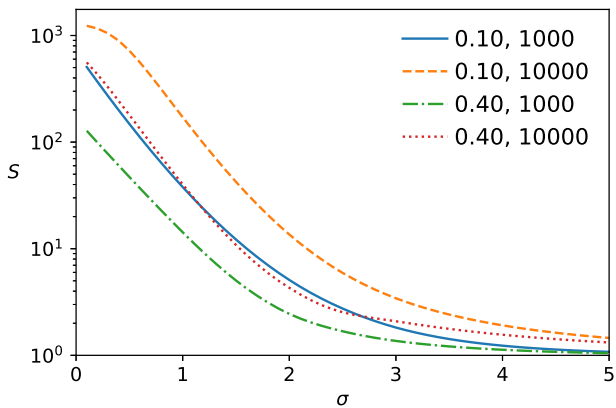}
\caption{Slope between the calculated p80s using $f={\rm sum}$ vs using $f={\max}$, for $\sigma_1=3$ and the $(q,n)$ indicated in the legend. }
\label{fig3}
\end{figure}

If we set activity durations to zero we can investigate $\max$ vs ${\rm sum}$ without the additional complication of free floats. Given the log-normal $p(h)\sim e^{-[log(h/\mu)]^2/2\sigma^2}/h$, we expect $S\rightarrow1$ when $\sigma\rightarrow\infty$. This asymptotic behavior is corroborated in Fig. \ref{fig1}.
The same behavior is obtained when the activity durations follow a sub-exponential distribution as well. For example, a log-normal distribution $p(d)\sim e^{-[log(d/\mu_1)]^2/2\sigma_1^2}/d$. I will set $\mu_1=1$ since $\sigma_1$ controls the shape of the distribution tail. Setting $\sigma_1=1$ and $\sigma_1=3$ we obtain the plots in Figs. \ref{fig2} and \ref{fig3} respectively. The asymptotic behavior $S\rightarrow1$ when $\sigma\rightarrow\infty$ is corroborated.

Regarding the network parameters, for a given number of activities, $S$ is closer to 1 for $q=0.4$ than $q=0.1$. The networks with larger duplication index $q$ have a broader degree distribution and a smaller diameter. In contrast, networks with small $q$ tend to be closer to a linear chain. This observation suggests that the distinction between using $f={\max}$ or $f={\rm sum}$ is less relevant in complex networks with wide degree distributions and smaller diameter.  

\begin{figure}[t]
\includegraphics[width=3.3in]{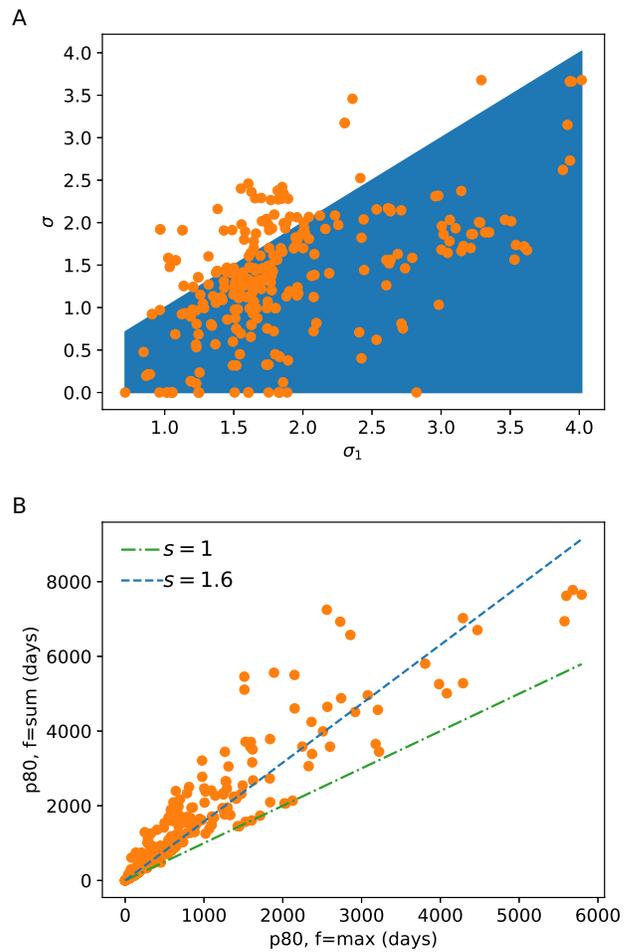}
\caption{A) Plot of $\sigma$ vs $\sigma_1$ for construction projects. Each symbol represents a construction project schedule. The shaded background highlights the area where $\sigma<\sigma_1$. B) $f={\rm sum}$ vs $f=\max$ p80 forecast for real construction projects (circles). The dashed line represents equal values. The dashed-dotted line represents the best linear fit through the origin. The activity network and activity durations are from the real projects. The parameters of $p(h)\sim e^{-[log(h/\mu)]^2/2\sigma^2}/h$ are project specific and inferred from observed delays for finished activities (see $\sigma$ in panel A) .}
\label{fig4}
\end{figure}

We have estimated $\sigma$ and $\sigma_1$ for construction projects in the Nodes \& Links database. For each project schedule in the warehouse we estimated $\sigma$ as the variance of the logarithm of reported delays (actual finish date $-$ planned finish date). We estimated $\sigma_1$ as the variance of the logarithm of activity durations. The delay variance $\sigma$ has an increasing trend with increasing the duration variance $\sigma_1$. Furthermore, there are several projects with low and high delay variance $\sigma$. To test the max bound to the sum, we carried on Monte Carlo simulations on top of real construction projects (\ref{fig4}B). The slope of the forecasted p80s using $f={\rm sum}$ vs $f=\max$ is 1.6, in the range of what observed for synthetic networks in the region $\sigma>2$ (Figs. \ref{fig1})-\ref{fig3}).

If using $f=\max(x,y)$ gives similar results as using $f={\rm sum}$ then we can choose either. As shown below, using  $f=\max$ has some advantages. For $f=\max$ we can solve Eq. (\ref{forward_delay}) iteratively
\begin{equation}
z_{i,t+1} = \max\left\{ \max_{j\in I_i}\left[ \max(0,w_{ij}+z_{j,t})\right] h_i\right\}\ ,
\label{forward_delay_iter} 
\end{equation}
Defining $w_{ii}=0$ we rewrite this equation as
\begin{equation}
z_{i,t+1} = \max_{j|j\in I_i\cup\{i\}} (w_{ij}+z_{j,t})\ .
\label{maxapp}
\end{equation}
Replacing $\max$ by $\oplus$ and $+$ by $\otimes$ the tropical algebra becomes evident. The tropical algebra is defined by the semiring $(\mathbf{R}\cup\{-\infty\},\oplus,\otimes)$ with the sum operation defined as
$x\oplus y = \max(x,y)$ and the product as $x\otimes y = x+y$ \cite{butkovic10}. The tropical algebra has been used to investigate networks with cycles \cite{hei14}. For example, of transportation networks, where it is desired to have a route back and forward between any two nodes. The tropical algebra has been applied to many other systems that can be mapped to event networks, including mRNA translation by ribosomes \cite{brackley12} and the  Conway’s Game of Life \cite{sakata20}. Using the tropical algebra we rewrite  (\ref{maxapp}) as
\begin{equation}
\vec{z}_{t+1} = L\otimes \vec{z}_t\ ,
\label{maxplus} 
\end{equation}
where
\begin{equation}
L = \left\{
\begin{array}{ll}
0 & {\rm if}\ i=j\\
w_{ij} & {\rm if}\ j\in I_i\\
-\infty & {\rm otherwise}
\end{array}\right.
\label{L}
\end{equation}
is the local weights matrix. $L$ encodes the free floats of direct relations, $L^{\otimes 2}$  the maximum free float sum of  paths up to length 2 and $L^{\otimes l}$  the maximum free float sum of paths up to length $l$. Since
project networks are represented by directed acyclic graphs, there is no path larger than the graph diameter $D$. Therefore $L^{\otimes l}=L^{\otimes D}$ for $l\geq D$. After $D$ iterations starting from $\vec{z}_0=\vec{h}$ the system will reach the steady state solution
\begin{equation}
\vec{z}_{\infty} = T\otimes \vec{h}\ ,
\label{maxplus_oo} 
\end{equation}
where
\begin{equation}
T = L^{\otimes D}\ ,
\label{T}
\end{equation}
is the total float matrix. The element $-T_{ij}$ equals the maximum $h_j$ that yields $z_{i,\infty}=0$.

Equation (\ref{maxplus_oo}) represents a principle of statistical independence of exogenous delays, in the tropical algebra sense. Each exogenous delay contributes independently of the others to the final propagated delay. This property is a direct consequence of the $f=\max$ merging function. Calculating the cumulative distribution function for the delay at each activity is now straightforward. Since exogenous delays are independent we calculate the maximum among the delay cascades propagated from each exogenous delay.  If $F_i(h)={\rm Pr}(h_i\leq h)$, $G_i(z) = {\rm Pr}(z_{i,\infty}\leq z)$ and $x_i$ is the early start date, then from Eq. (\ref{maxplus_oo}) we obtain the probability distributions for the activity end dates
\begin{equation}
G_i(y) = \prod_{j} F_j(y-x_i-T_{ij})\ .
\label{CDF}
\end{equation}
This equation represents an analytical approximation to the end date distribution of all activities in a project network. In this equation we have a clear separation between the influence of the distributions of exogenous delays $F_j(h)$, the planned end date $x_i$ and the activity network properties encoded in the total floats $T_{ij}$. From there we can calculate other relevant quantities. The probability to finish on time $G_i(0)$ and the end date $y_i(p)$ with $p$ confidence $G_i[y_i(p)]=p$. We can determine the impact of accelerating activity execution to catchup with delays as well, by taking into account that $T_{ij}=T_{ij}(\vec{d})$.

The non-trivial region where $f=\max$ is a good approximation to $f={\rm sum}$ remains to be explained. Intuitively, if both the delays and the free floats have sub-exponential distributions, and we extract a delay and a free float from those distributions, then one of them will be much larger than the other with high probability. In that case, the delays that exceed the free floats will be large and $f=\max$ is a good approximation for $f={\rm sum}$. More work is required to translate this intuition into analytical demonstrations. Furthermore, a delay in the activity start date could increase the delay in the activity finish date, above what expected if the activity start wasn't delayed (supper-additive). If the rate per activity of such events is large then we should expect a breakdown of the tropical approximation

In conclusion, I have obtained a tropical approximation to the end date distribution of activity networks. This approximation replaces ${\rm sum}$ by $\max$ in the algebraic equations that merge endogenous with exogenous delays. The tropical approximation is close to simulated data when (i) the distribution of exogenous delays belongs to the sub-exponential family and (ii) the variance of the exogenous delays is larger than 1. We found that the tropical approximation is a good bound when tested in the Nodes \& Links construction project database.

I thank Christos Ellinas and Georgios Kalogridis for manuscript comments. This work was partly supported by European Union's Horizon 2020 and the Cyprus Research \& Innovation Foundation under the SEED program (grant agreement 0719(B)/0124). Nodes \& Links provided support in the form of salary for AV, but did not have any additional role in the study design, data collection and analysis, decision to publish, or preparation of the manuscript. 

\bibliographystyle{apsrev4-1}


\bibliography{maxplus_corrected.bbl}

\end{document}